\documentclass[fleqn,usenatbib]{mnras}

\usepackage{newtxtext,newtxmath}
\usepackage[T1]{fontenc}

\DeclareRobustCommand{\VAN}[3]{#2}
\let\VANthebibliography\thebibliography
\def\thebibliography{\DeclareRobustCommand{\VAN}[3]{##3}\VANthebibliography}

\usepackage{xcolor}
\usepackage{float}

\usepackage{graphicx}	% Including figure files
\usepackage{amsmath}	% Advanced maths commands

\usepackage{amssymb}	% Extra maths symbols

\title[GRBs lensed by clusters]{An archival search for gamma-ray bursts gravitationally lensed by galaxy clusters}

\author[D. Ryczanowski et al.]{
Dan Ryczanowski,$\!^{1,2\thanks{E-mail: dan.ryczanowski@port.ac.uk}}$
Benjamin P. Jones,$\!^{2\thanks{Current address: Met Office, FitzRoy Road, Exeter, Devon, EX1 3PB, UK}}$
Benjamin P. Gompertz,$\!^{2,3}$
Graham P. Smith$^{2,4}$
\\
\\
$^{1}$Institute of Cosmology and Gravitation, University of Portsmouth, Burnaby Road, Portsmouth, PO1 3FX, UK\\
$^{2}$School of Physics and Astronomy, University of Birmingham, Edgbaston, Birmingham, B15 2TT, UK\\
$^{3}$Institute for Gravitational Wave Astronomy, University of Birmingham, Edgbaston, Birmingham, B15 2TT, UK\\
$^4$Department of Astrophysics, University of Vienna, T\"urkenschanzstrasse 17, 1180 Vienna, Austria\\
}

\pubyear{2026}

\begin{document}
\label{firstpage}
\pagerange{\pageref{firstpage}--\pageref{lastpage}}
\maketitle

%%%%%%%%%%%%%%%%%%%%%%%%%%%%%%%%%%%%%%%%%%%%%%%%%%
\begin{abstract}
Discoveries of gamma-ray bursts (GRBs) have become commonplace in recent decades, totalling $\mathcal{O}(10^4)$ unique detections across various missions. However, there have been no confirmed discoveries of a gravitationally-lensed GRB, despite expected lensing rates of $\sim1$ in $10^{3}$. In light of this, we complete an archival search for lensed GRBs by cross-matching well-localised \emph{Swift}/XRT-detected bursts with a large all-sky sample of galaxy clusters as potential lenses. We find a total of 17 candidate lensed GRBs defined by a 2 arcminute search radius from a cluster in our sample. 14 of our candidates are either confirmed to be at higher redshifts than their cross-matched cluster, or are consistent with a higher redshift origin based on the Amati relation between $E_{p,i}$ and $E_{\rm iso}$ of GRBs, indicating they are, at some level, lensed by their nearby cluster. Using the Amati relation and the lens-GRB separation, we quantify the magnification experienced by each GRB. We find $\mu < 10$ for all except for one candidate, GRB~071031, which is consistent with $\mu > 10$, but is uncertain. Another candidate, GRB~050509B, does not have a directly measured redshift, but was previously assumed to be at the redshift of its nearby cluster, $z=0.225$. We produce a lens model of this cluster and show that GRB~050509B is consistent with $z>1$ and magnified by $\mu\simeq2-6$. We present these findings in anticipation of future lensed GRB discoveries enabled by facilities such as the Vera C. Rubin Observatory in the coming years.
\end{abstract}


%%%%%%%%%%%%%%%%% BODY OF PAPER %%%%%%%%%%%%%%%%%%

\section{Introduction}
The study of gravitationally lensed explosive transients is a rapidly evolving field, offering unique approaches to address open questions in fundamental physics, cosmology, and astrophysics \cite[and references therein]{Oguri19,Grespan2023,Smith2025}. Despite significant efforts made across numerous different transient sources, the confirmed lensed explosive transient detections to date have been exclusively supernovae, including multiply-imaged events \citep{Kelly15,Goobar17,Rodney21,Kelly22,Chen22,Goobar23,Frye24,Pierel24,Taubenberger25,Johansson25}, and magnified singly-imaged events \citep[e.g.][]{Goobar09,Amanullah11,Patel14,Rodney15,Rubin18}. With lensed supernova detections set to increase by at least an order of magnitude thanks to the Vera C. Rubin observatory's Legacy Survey of Space and Time \citep[e.g.][]{Goldstein2019,Arendse23,SainzdeMurieta23,PontePerez2025}, it is opportune to consider the discovery of other explosive transients, such as gravitationally lensed gamma-ray bursts (GRBs).

Interest in gravitationally lensed GRBs dates back at least as far as the work that firmly established GRBs as extragalactic sources \citep{Paczynski1986, Porciani2001}. GRBs are very luminous events that are associated with the core collapse of massive stars and binary compact object mergers, and are typically detected at redshift $z\simeq2$ \citep{Frail2001,Berger2003,Jakobsson2006,Coward_2013,Burns2023}. Their typical redshifts are therefore well-matched to being lensed by foreground galaxies and clusters of galaxies located along the line of sight at redshifts of $z\simeq0.2-0.5$. They are also, given the sub-second timing accuracy of GRB instruments and temporal structure of GRB light curves, well-matched to the discovery of small lenses with masses comparable to, e.g., intermediate mass black holes \citep{Mao1992, Grossman1994}.

Most searches for lensed GRBs have so far focused on micro/milli-lensing \citep{Masetti2003, Perley2007, Rapoport2012, Paynter2021, Kalantari2021, Veres2021, Wang2021, Roberts2022, Chen2022, Mukherjee2024a, Mukherjee2024b}, whilst a few have explored strong lensing by galaxies and clusters \citep{Li&Li14,Ahlgren20}. None of these studies have uncovered irrefutable evidence for the discovery of a lensed GRB, indicating several challenges are yet to be overcome \citep{Levan2025}. The overarching problem for micro/milli-lensed cases is breaking the degeneracy between lensing and non-lensing interpretations of structure within individual GRB light curves. For strongly-lensed cases, the problem is being able to correctly interpret multiple GRB detections as being separate unrelated GRBs, or lensed images of a single source. A direct path to break the degeneracy is to obtain independent evidence of a lens at the location of the GRB detection(s). However, this is challenging because GRB prompt emission is typically localised to sky regions of $\Omega\gtrsim1000\,\rm deg^2$, within which many lenses of all scales reside \citep[and references therein]{Shajib2024}, and just $\simeq20\%$ of GRB detections are localized to the scale of a strong lens via their afterglow emission \citep[and references therein]{Levan2025}. Further complications arise when considering strong lensing of GRBs in light of uncertainties on how similar two lensed images of a GRB will be, given that each image may also be micro-lensed \citep{Williams1997,Perna2009} and due to open questions on the structure of GRB jets \citep{Levan2025}, which may cause individual lensed images to be non-identical. 

We take a new approach to analysing historic GRB detections in search of signatures of gravitational lensing, in part motivated by forecasts that $\simeq10-20$ of the $\simeq10^4$ GRB detections to date were gravitationally magnified by a factor of at least $\mu\gtrsim3$ \citep{Smith2025}. We cross-match the subset of GRB detections that have sky localisation uncertainties of $\sim1\,\rm arcsec$ from \textit{Swift} afterglow detections with large catalogues of groups and clusters of galaxies. These groups and clusters are capable of lensing a distant source, even if they have not yet been identified as lenses \citep{Ryczanowski20}. We therefore sidestep the poor localisations of prompt GRB emission and the physical uncertainties that currently impact cross-matching with most GRBs. Using the results of this cross-match, we take a refreshed look at whether lensed GRBs have already been detected, as a precursor to anticipated future discoveries from prompt optical follow-up of candidate lensed GRBs with wide field-of-view instruments, such as the La Silla Schmidt Telescope (\citealt{Miller2025}), the Gravitational-wave Optical Transient Observer (GOTO; \citealt{Steeghs2022}) and the Vera C.\ Rubin Observatory's Simonyi Survey Telescope \citep[a.k.a. Rubin,][]{Andreoni2024}.

The structure of the paper is as follows: in \autoref{sec:data}, we discuss the data used -- including the GRB detections and the groups and clusters that make up our sample of capable lenses. In \autoref{sec:analysis} we describe in detail the cross-matching between the GRB and cluster samples that forms the basis of our lensed GRB search. We provide a more detailed analysis of our strongest candidate in \autoref{sec:GRB050509B}, and finally discuss our results in \autoref{sec:discussion}. All magnitudes quoted are in the AB system. 

\section{Data}
\label{sec:data}

\subsection{GRB sample}

Confident matching of transients to potential lenses -- especially where measurements of the gravitational magnification factors are desired -- requires localisation precisions at the arcsecond level. We therefore restrict our GRB sample to bursts discovered by the \emph{Neil Gehrels Swift Satellite} \citep[\emph{Swift};][]{Gehrels04} with detections from its X-ray Telescope \citep[XRT;][]{Burrows04} which can meet the $\sim1$ arcsec precision requirement. While this choice significantly restricts the number of available GRBs compared to the full sample discovered by all-sky monitors like the \emph{Fermi} Gamma-ray Burst Monitor \citep[GBM;][]{Meegan09}, it greatly reduces the number of false positive matches that would be expected with the $\gtrsim 10$ square degree localisation regions produced by GBM. 

The version of the \textit{Swift}/XRT catalogue used in this study was accessed on 10 June 2024 and contains 1336 GRBs. A sample of this size is statistically significant for identifying gravitationally lensed GRBs, given that the ratio of lensed GRB detections to the number of un-lensed GRB detections is forecast to be $\mathcal{\rho}\simeq10^{-3}$ \cite{Smith2025}.

\subsection{The cluster sample}
\label{sec:cluster_sample}
We identify clusters as the choice of lenses for cross-matching in our lensed GRB search. The population of  ``confirmed'' lenses (those with identified arcs or lensing features) is incomplete, and therefore, focusing on confirmed lenses would greatly restrict the search. The alternative is to consider ``capable'' lenses; objects such as galaxy clusters and luminous red galaxies (LRGs) which are common among known lenses, but do not show any known lensing features \citep{Ryczanowski20}. Capable lenses are significantly more numerous than confirmed lenses, but ensure a much more complete sample. However, galaxy-scale objects are incredibly numerous, for example, the recent DESI Legacy Survey DR10 release contains over 2.5 billion such objects in approximately half of the sky \citep{Dey2019}. Therefore focusing on cluster-scale objects helps to strike a balance between completeness of lenses and false positive lensed GRB candidates, whilst still accounting for roughly half of the halo mass-integrated strong lensing optical depth for sources at the mean GRB redshift of $z=2$ \citep{Robertson20}.

The stochastic but homogeneous distribution of GRB detections requires a cluster catalogue that is both all-sky and sufficiently deep to include all objects that are efficient lenses of the high redshift GRB population. No single catalogue optimally fulfils these requirements, due to the trade off between depth and sky coverage, so we opt to use a variety of catalogues, each constructed using independent methods. We describe those utilised for our cross-match in the remainder of this section.

\subsubsection{CALICO}
The Catalogue of All-sky Infrared Cluster Overdensities (CALICO) is a collection of high density lines of sight of red early type galaxies, which form the basis of most known lenses on both galaxy and cluster scales. Constructed with the purpose of discovering lensed transients within wide-field surveys, CALICO provides regions of sky where many capable lenses are found, significantly reducing the relevant search area for lensed transients. We hereafter refer to entries in CALICO as ``clusters'', despite the catalogue containing objects on both group and cluster scales, as well as high surface density regions of red objects at multiple redshifts. 

CALICO is built based on the method presented in \citet{Ryczanowski22}, which we briefly summarise here. As the name suggests, it is an all-sky catalogue, and contains candidate clusters up to $z\sim1$. Clusters are selected by identifying high density regions of infra-red galaxies from two hemispherical surveys: UHS \citep[UKIRT Hemisphere Survey,][]{Dye18} which covers the north and VHS \citep[Vista Hemisphere Survey,][]{McMahon13} covering the south. A discretised grid of galaxy positions from these surveys is taken as input, which is then convolved with a difference-of-Gaussians kernel to produce a smooth 2D density map of peaks and troughs representing over- and under-dense regions. The pixels of these resulting density maps are then normalised to the values obtained by performing the same process on a pixel map of an identical number of randomly distributed galaxies (i.e. a map with no clustering). Therefore, the normalised map represents how densely populated a region of sky is compared to pure randomness. Numerically, the value of each pixel in the normalised map represents a so-called effective signal-to-noise ratio (SNR$_{\rm eff}$), which is used to asses the significance of each overdensity. A cut on SNR$_{\rm eff}$ is used to select peaks in the normalised maps that translate to real and significant overdensities; the version of CALICO used in this study cuts at SNR$_{\rm eff} > 5$, resulting in a total of 118,152 clusters.

A key motivation of CALICO is to supplement existing cluster catalogues that are restricted in sky coverage, and to extend on all-sky catalogues that are based on shallower data. It is important to note that although CALICO is purpose-built it is not a direct replacement for other catalogues. The other cluster catalogues in this section infer additional information such as the cluster mass and redshift (or proxies for these quantities). The cluster redshift can be used during cross-matching to compare with the GRB redshift (if known) to see if it is likely to have been lensed by the cluster, or is instead located within or in front of the cluster. CALICO does not infer a redshift for its clusters since it uses only a single photometric band, however public photometric redshift catalogues e.g. from DESI Legacy Survey \citep{Dey2019} can often be used to estimate a cluster redshift.

\subsubsection{WHY Catalogue}
The \citet*[][]{WHY18}, hereafter, WHY cluster catalogue, is an all-sky catalogue containing 47600 clusters, located by searching for galaxy overdensities in both angular and photometric redshift space around candidate LRGs that typically lie within cluster cores. WHY utilises galaxy photometry from three all-sky catalogues: the Two Micron All Sky Survey (2MASS), the Wide-field Infrared Survey Explorer (WISE), and SuperCOSMOS. Objects from these surveys, and hence the clusters in WHY, are at the lower redshift end ($z\leq 0.3$) of the population of efficient lenses of GRBs which will extend close to their typical source redshift $z\sim 2$. This underlines the need to supplement with higher redshift clusters from other catalogues. Clusters in WHY have also been cross-matched with detections of extended x-ray sources from {\it ROSAT} and {\it XMM-Newton}. Thus, although we do not explicitly use any X-ray cluster catalogues in our sample, this ensures they have some representation.

\subsubsection{redMaPPer}
The redMaPPer algorithm \citep{Rykoff14} locates galaxy clusters using the red sequence technique, which detects excesses of galaxies with consistent colours. The similarity of their colours pertains to formation within the same environment and at the same cosmic time, and is often visualised by a linear fit to points on a colour-magnitude diagram. redMaPPer has been applied to the Sloan Digital Sky Survey data release 8 (SDSS DR8) to create a catalogue of about 25,000 clusters over $\sim 10,000$ square degrees, out to a redshift of $z\sim0.55$. Compared to the WHY catalogue, it has the advantage of reaching a higher redshift, but at the cost of being restricted to the SDSS footprint. The algorithm also estimates the photometric redshift and richness of clusters using statistical methods, which are useful in this analysis to determine whether a cross-matched GRB is likely to be located in front, within or behind the matched galaxy cluster.

\subsubsection{Other cluster catalogues}

We cross-match against a set of clusters detected through the distortion of the cosmic microwave background by hot intra-cluster gas -- the Sunyaev-Zeldovich effect. We utilise both the Planck Sunyaev-Zeldovich 2 \citep[PSZ2,][]{Planck16} and Atacama Cosmology Telescope \citep[ACT,][]{Hilton21} cluster catalogues in our sample. PSZ2 covers the entire sky, whilst ACT covers $\sim 13,000$ square degrees. However, the limited coverage of ACT is compensated by a greater level of depth and completeness than is achieved in PSZ2. Because the typical positional uncertainty for a cluster in PSZ2 ($\gtrsim 1$ arcmin) is larger than the average strong lensing region of a typical galaxy cluster ($\theta_E \sim$ 20 arcsec), we instead use coordinates for these clusters centred on likely bright central galaxies (BCGs) where available \cite{JSmith2023}, instead of the coordinates provided in PSZ2. This ensures our search targets the centre of mass, and hence the strong lensing region of these diffuse clusters.

We additionally make use of a catalogue of 130 spectroscopically-confirmed cluster strong lenses from the literature, assembled by \citet{Smith2018} and originating from a variety of original sources. Many of these are well studied and have detailed lens models available.

\subsubsection{Duplicates in the cluster catalogues}
The independent nature of the methods used to produce the various cluster catalogues means there is no protection against duplications in our cluster sample. Even though methods differ, the underlying reliance on photometric catalogues in regions with vast overlapping footprints almost guarantees some duplication will occur. To quantify this, we perform a cross-match of the coordinates in our sample with itself, and consider any sets of objects separated by less than 1 arcminute to be the same cluster. The 1 arcminute separation is chosen as it corresponds approximately to the uncertainties on cluster positions within CALICO, the largest catalogue within our sample. This self cross-matching finds that out of 197,054 clusters from the various catalogues, there are 20,635 duplicates, leaving us with a total of 176,419 unique clusters in our sample. Due to duplicates being only at the $\sim10$ per cent level, and since coordinates of cluster centres may differ between catalogues, we did not consider duplicates until after cross-matching and only for those with a nearby GRB, which were reviewed on a case-by-case basis.

\section{Cross-match analysis and results}
\label{sec:analysis}

\subsection{Cross-matching the \textit{Swift}/XRT catalogue and cluster sample}

We cross-match the 1336 GRBs from the \textit{Swift}/XRT catalogue to the 176,419 clusters in our sample, and compute the on-sky separation angle, $\theta_{\text{sep}}$. Our resulting matches are listed in \autoref{tab:matches}, where we show all GRB-cluster pairs with a $\theta_{\text{sep}} < 2'$. 2 arcminutes is chosen as a generous upper bound on the separation between a lensing cluster and the apparent position of an associated lensed source. This is based on models of the largest cluster from the Hubble Space Telescope Frontier Fields programme, MACS J0717.5+3745 \citep[HST FF;][]{Lotz_2017}, which have modelled\footnote{https://archive.stsci.edu/prepds/frontier/lensmodels/} critical curves with a semi-major axis of $\sim2'$ for a source at the mean GRB redshift of $z=2$, justifying $2'$ as the maximum search radius.

\begin{table*}
    \centering
    \caption{Cluster sample and \textit{Swift}-XRT GRB cross-matches within $\theta_{\text{sep}}\leq 2\,'$, where $\theta_{\text{sep}}$ is the on-sky separation in arcminutes between the GRB-cluster pair, ordered by  $\theta_{\text{sep}}$. Coordinates and redshifts are provided for each object in the matched pair, where GRB coordinates have positional error at the $90\%$ confidence level. GRB redshifts (where known) are all spectroscopic, besides GRB~050509B, which has previously been assumed to lie at the cluster redshift (see \autoref{sec:GRBs}). Cluster IDs marked with * are found in multiple catalogues, but we only quote one for brevity. Cluster redshifts (where given) are photometric; unless marked with $\dag$ for spectroscopic. For an explanation of the CALICO redshifts, and why some have two values quoted, see Appendix \ref{sec:calico_redshifts}. $^{a}$050509B redshift is not independently confirmed, and is assumed to be associated with the intervening cluster \citep{Bloom06}.}
    \label{tab:matches}
    \begin{tabular}{lllllllll}
        \hline
        GRB ID  & GRB $\alpha$,$\delta$ (J2000) & Err$_{90}$ $(\arcsec)$ & $z_{\text{GRB}}$ & Galaxy Cluster ID                      & $z_{\text{Cl}}$  & Cluster $\alpha$, $\delta$ (J2000) & $\theta_{\text{sep}}(\arcmin)$ \\ \hline
        091029  & 60.1776, -55.9554                    & 1.4                    & 2.75          & CALICO\_S19885                         & 0.31, 0.5                    & 60.1845, -55.9505                         & 0.3749                         \\
        191101A & 251.8372, +43.7407                   & 1.4                    & -             & redMaPPer 28493                        & 0.4052                   & 251.8460, +43.7438                        & 0.4238                         \\           
        180204A & 330.1333, +30.8378                   & 1.4                    & -             & CALICO\_N41307                         & -                        & 330.1250, 30.8307                         & 0.5983                         \\
        070419B & 315.7075, -31.2636                   & 1.5                    & 1.9588        & WHY J210250.8-311451                   & 0.1127                   & 315.7115, -31.2475                        & 0.9866                         \\
        071031  & 6.4058, -58.05919                    & 1.5                    & 2.69          & CALICO\_S41090                         & 0.55                  & 6.4309, -58.0693                          & 1.0015                         \\
        060712  & 184.0678, +35.5383                   & 1.6                    & -             & WHY J121621.2+353240*                  & 0.2876                   & 184.0881, +35.5446                        & 1.0756                         \\
        180614A & 3.0784, +46.9530                     & 1.7                    & -             & WHY J001222.4+465608                   & 0.3074                   & 3.0933, +46.9355                          & 1.2160                         \\
        131229A & 85.2317, -4.3963                     & 1.4                    & -             & CALICO\_S11700                         & -                    & 85.2346, -4.4235                      & 1.3010                         \\
        200215A & 34.0794, +12.7710                    & 1.4                    & -             & CALICO\_N856                           & 0.15, 0.34                   & 34.0742, +12.7495                       & 1.3277                         \\ 
        231230A & 245.2149, 58.1237                    & 2.0                    & -             & WHY J162058.7+580628                   & 0.1636                   & 245.24460, 58.10770                       & 1.3460                        \\
        081025  & 245.3667, +60.4756                   & 5.2                    & -             & WHY J162130.9+602713                   & 0.3098                   & 245.3790, +60.4537                        & 1.3634                         \\
        060306  & 41.0953, -2.1486                     & 1.4                    & 1.55          & WHY J024419.1-020749                   & 0.2658                   & 41.0796, -2.1304                          & 1.4398                         \\
        081211B & 168.2645, 53.8297                    & 2.1                    & -             & redMaPPer 4938* & 0.2156$^{\dag}$ & 168.2231, +53.8304   & 1.4730                \\
        210514A & 2.9618, -21.8945                     & 2.0                    & -             & CALICO\_S39568                         & 0.05, 0.45                   & 2.9444, -21.9110                          & 1.5530                         \\
        050509B & 189.0574, +28.9843                   & 5.4                    & 0.225$^{a}$         & redMaPPer 11161* & 0.2433         &  189.0877, +28.9915   & 1.6475                \\
        081128  & 20.8047, +38.1275                    & 1.6                    & -             & WHY J012306.3+380631                   & 0.3342                   & 20.7762, +38.1086                         & 1.7602                         \\
        090113  & 32.0573, +33.4287                    & 1.4                    & -             & WHY J020812.4+332352                   & 0.2733                   & 32.0518, +33.3977                         & 1.8775                         \\
        \hline
    \end{tabular}
\end{table*}

We find 17 GRB-cluster pairs with $\theta_{\rm sep} \leq 2'$; these are shown in Table~\ref{tab:matches}, ordered by $\theta_{\rm sep}$. The GRB properties are listed in Table~\ref{tab:GRBs}. 4/17 have independently determined redshifts that place them behind their matched clusters: GRB~060306 \citep[$z = 1.55$;][]{Perley13}, GRB~070419B \citep[$z = 1.9588$;][]{Kruhler12}, GRB~071031 \citep[$z = 2.69$;][]{Fox08} and GRB~091029 \citep[$z = 2.75$;][]{Chornock09}. GRB~050509B does not have an independently measured redshift, but its proximity to its cross-matched cluster, ZwCl\,1234.0$+$02916 (redMaPPer 11161 in our sample) has been previously noted, and often adopts its redshift of $z = 0.225$ \citep{Bloom06}. Since this association is not confirmed, there is a possibility that GRB~050509B is instead behind the cluster and hence lensed by it -- this was previously investigated by \citet{Pedersen05}, and we discuss their findings further in the context of our own in \autoref{sec:GRB050509B}.

\subsection{Gamma-ray burst analysis}\label{sec:GRBs}

\begin{figure}
    \centering
    \includegraphics[width=\columnwidth]{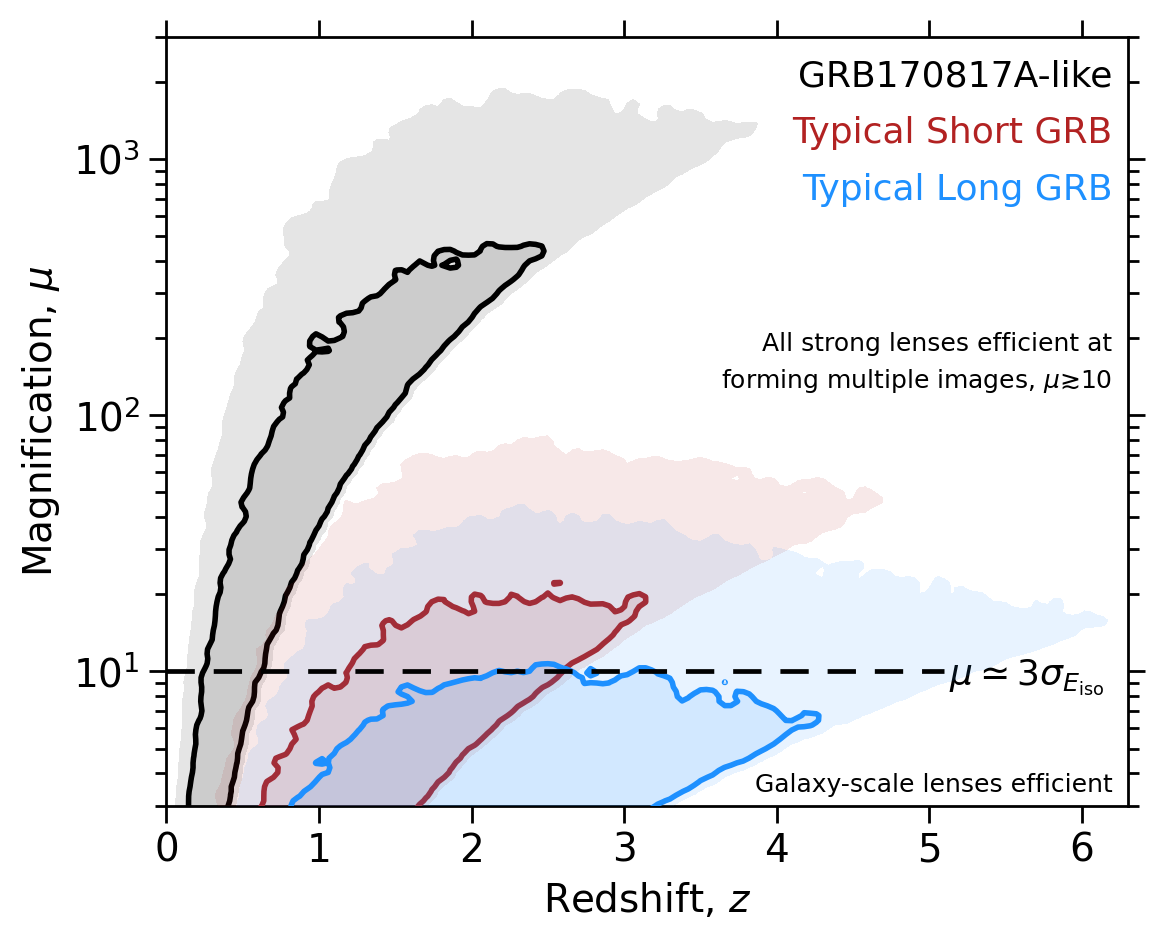}
    \caption{Forecast location of detectable gravitationally-lensed GRBs in the magnification-redshift plane, based on the multi-messenger model presented by \citet{Smith2025}, showing long and short GRBs (blue/lower and red/middle respectively), plus short GRBs viewed at a similar off-axis angle as GRB~170817A (black/upper), for comparison. Shaded regions cover 99\% of the forecast detectable lensed populations, and each bold contour encloses 90\% of these detectable objects. The horizontal scatter in the $E_{p,i}-E_{\rm iso}$ relation is approximately an order of magnitude (see \autoref{fig:Ep_Eiso}), so sources magnified by a factor $\mu \gtrsim 10$ (horizontal dashed line) are expected to be outliers from the relation at $\gtrsim 3\sigma_{E_{\rm iso}}$. The $\mu=10$ line also represents the level of gravitational magnification below which cluster-scale gravitational lenses are inefficient at forming multiple images \citep[e.g.][]{Fox2022,Smith2023}, highlighting that almost all of the detectable lensed populations of long GRBs, and $\sim2/3$ of short GRBs will be singly-imaged.}
    \label{fig:muredshift}
\end{figure}

\begin{figure*}
    \centering
    \includegraphics[width=0.65\textwidth]{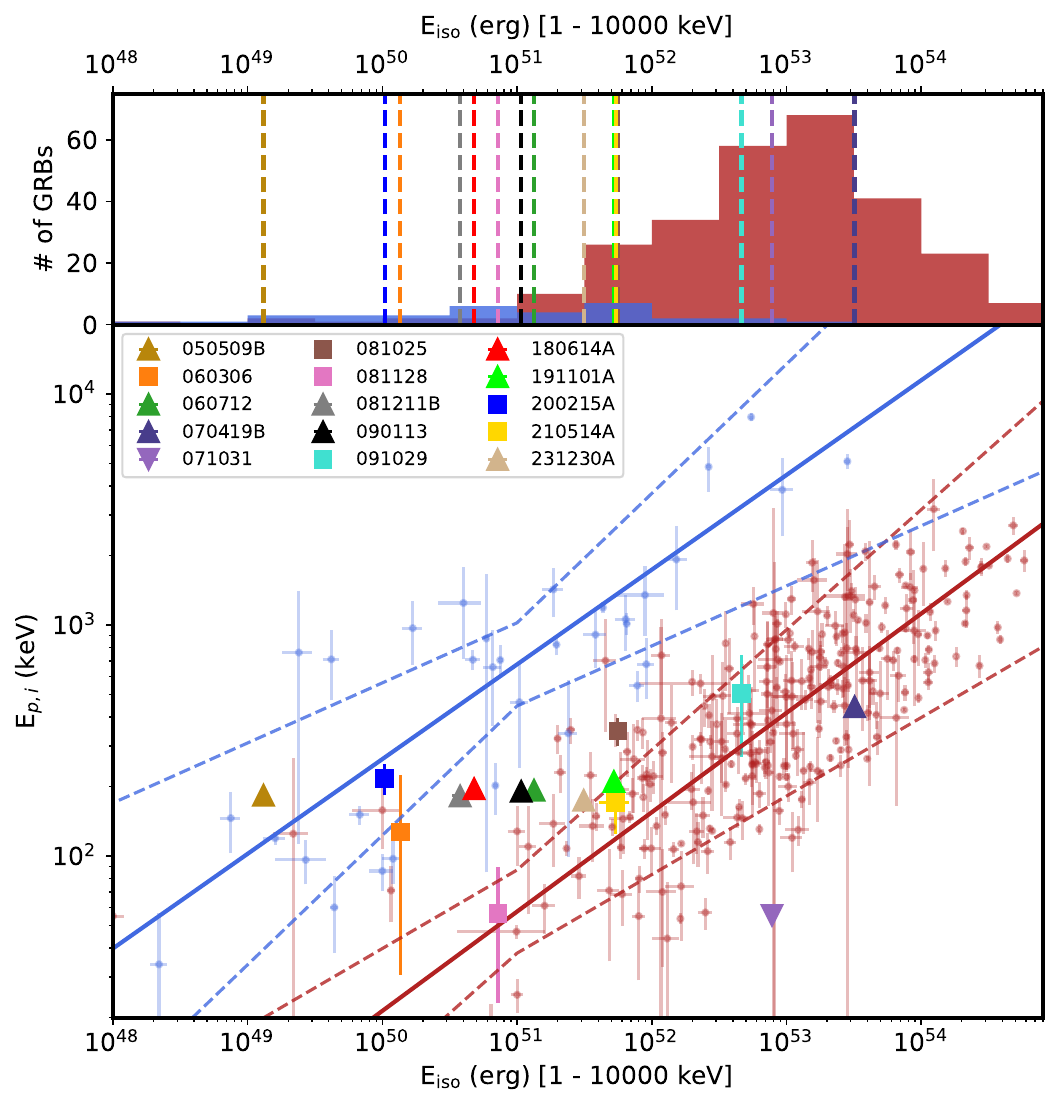}
    \caption{The $E_{p,i}$ -- $E_{\rm iso}$ relations for long GRBs (red) and short GRBs (blue), plotted alongside the \citet{Minaev20} sample, which is additionally represented in the top histogram. Dashed lines on the scatter plot mark the $3\sigma$ confidence intervals for the relations. The 15/17 GRBs in our sample with redshifts measured for either the GRB or the matched cluster are shown by the large markers, and by dashed vertical lines of corresponding colour on the top histogram. Square symbols represent GRBs with a measured $E_p$, upright triangles indicate lower limits of $E_{p,i} > (1+z)150$\,keV and inverted triangles indicate upper limits of $E_{p,i} < (1+z)15$\,keV. For GRBs with known redshifts, we expect the lensed population to be bright outliers from the relation in $E_{\rm iso}$, as long as their magnification is greater than the scatter in $E_{\rm iso}$, approximately $\mu \sim 10$ (\autoref{fig:muredshift}). We assume the cluster redshift (\autoref{tab:matches}) for GRBs with no independent redshift measurement. This assumption leads to the majority of this subsample appearing to be significantly under-luminous, strongly suggesting a higher redshift origin than the nearby cluster, and hence some level of magnification will be experienced by these GRBs.}
    \label{fig:Ep_Eiso}
\end{figure*}

GRBs show a positive correlation between the rest-frame 
%redshift-corrected 
spectral peak of their prompt emission ($E_{p,i}$) and their isotropic equivalent energy release in gamma-rays \citep[$E_{\rm iso}$;][also known as the Amati relation]{Amati02,Amati06}. Detectable gravitationally-lensed GRBs of \emph{known redshift} will be gravitationally magnified, and thus their $E_{\rm iso}$ will be over-estimated if the presence of lensing is not identified. This may cause lensed GRBs to be bright outliers in the $E_{p,i}-E_{\rm iso}$ plane if no lensing (i.e. a magnification of $\mu=1$) is assumed, depending on how $\mu$ compares with the scatter on the $E_{p,i}-E_{\rm iso}$ relation. Conversely, detectable gravitationally-lensed GRBs of \emph{unknown redshift} will typically be faint outliers in the $E_{p,i}-E_{\rm iso}$ plane if they are assumed to be associated with the lens and hence incorrectly assigned the redshift of the lens. This is because the typical lens ($z\simeq0.3$) and GRB ($z\simeq2$) redshifts imply a factor $\simeq100$ systematic error in $E_{\rm iso}$ for such objects. Moreover, gravitational magnification is unlikely to compensate for this distance-related systematic for GRBs, because they are intrinsically bright and thus lensed GRBs are expected to be dominated by low magnification events with $\mu\simeq2-10$ \citep[Fig.~\ref{fig:muredshift} and][]{Smith2025}. 

We therefore investigate the location of the candidate lensed GRBs from our cross-match analysis in the $E_{p,i}-E_{\rm iso}$ plane. Placing a GRB in this plane requires a known (spectroscopically confirmed) or estimated redshift for each GRB. We therefore exclude GRB180204A and GRB131229A from this analysis because neither they nor their associated cluster have a known or estimated redshift (Table~\ref{tab:matches}; Appendix~\ref{sec:calico_redshifts}). This leaves a sample of 15 GRBs, all of which have a cluster redshift, $z_{\rm Cl}$, and 4 of which have a spectroscopic GRB redshift, $z_{\rm GRB}$. The 11 GRBs without $z_{\rm GRB}$ are assigned the same redshift as their cross-matched cluster for the purpose of this analysis. Those appearing as faint outliers would imply their true redshifts are larger than $z_{Cl}$, and hence will, at some level, be lensed by the cluster. All except one of the 15 GRBs are long GRBs, with GRB~050509B being the sole short GRB by virtue of having a measured duration of $t_{90} < 2$\,s \citep{Kouveliotou93}, where $t_{90}$ is the time elapsed during emission of the central 90 per cent of gamma ray counts.

To calculate the observer frame $E_{p}$ and measure the fluence of each GRB, we fit the time-averaged BAT spectral data over the $t_{90}$ interval. Data are downloaded from the UKSSDC and processed with the standard \emph{Swift} GRB pipeline {\sc batgrbproduct} v2.48. Processed data are then fit in {\sc xspec} v12.11.1 \citep{Arnaud96}. Each spectrum is fit with a single power-law model (PL), a power-law with an exponential cutoff (Cutoff) and the GRB continuum Band function \citep{Band93}, which features two power-law segments $\alpha$ and $\beta$ smoothly connected at a characteristic energy $E_c = E_p/(2-\alpha)$. We accept the value of $E_p$ for the GRB where we find a $> 3\sigma$ improvement in the statistical fit for the Cutoff or Band model compared to the PL model, as measured by an f-test. Where available, we supplement our results with published fits from the \emph{Fermi}-GBM catalogue \citep{Gruber14,vonKienlin14,Bhat16,vonKienlin20} and GCNs from \emph{Konus}-WIND, both of which have wider bandpasses than BAT and therefore offer a better chance of measuring $E_p$. In cases where a power-law fit is preferred, we can still impose limits on the location of $E_p$ from the photon index. Since the peak of the spectrum in $\nu F_{\nu}$ units occurs when $\alpha = 2$, we assume $E_p > 150$\,keV (the high end of the BAT bandpass) where $\alpha < 2$ and $E_p < 15$\,keV (the low end of the BAT bandpass) where $\alpha > 2$.

\begin{table*}
    \centering
    \caption{Properties and spectral fits for the 17 GRBs that lie within 2' of a galaxy cluster in our catalogue. $t_{90}$ values for \emph{Swift} bursts are taken from \citet{Lien16}. $^a$\emph{Fermi}-GBM catalogue. $^b$\citet{Lien16}. $^c$\citet{Poolakkil20}. $^d$\citet{Ridnaia21}. $^{\dagger}2\sigma$ improvement over the power-law fit.}
    \label{tab:GRBs}
    \begin{tabular}{ccccccccc}
    \hline
    \hline
    GRB & $t_{90}$ & Model & $\alpha$ & $\beta$ & $E_{p}$ & Fluence & $\chi^2$ & dof \\
     & (s) & & & & (keV) & (erg\,cm$^{-2}$) \\
     \hline
     050509B & $0.024 \pm 0.009$ & PL & $1.13 \pm 0.32$ & & $> 150$ & $(6.03 \pm 0.54) \times 10^{-9}$ & $52.8$ & 56 \\
     060306 & $60.9 \pm 3.4$ & Band$^{\dagger}$ & $0.75 \pm 0.67$ & $2.24 \pm 0.32$ & $49.9 \pm 37.9$ & $(2.95 \pm 0.06) \times10^{-8}$ & $45.1$ & 54 \\
     060712 & $30.9 \pm 6.6$ & PL & $1.75 \pm 0.19$ & & $> 150$ & $(1.29 \pm 0.06) \times10^{-6}$ & $55.0$ & 56 \\
     071031 & $180.0 \pm 10.0$ & PL & $2.28 \pm 0.18$ & & $< 15$ & $(7.72 \pm 0.31)\times10^{-7}$ & 46.39 & 56 \\
     070419B & $238.0 \pm 14.3$ & PL & $1.67 \pm 0.03$ & & $> 150$ & $(6.56 \pm 0.04) \times 10^{-6}$ & $32.2$ & 56 \\
     081025$^a$ & $22.8 \pm 1.0$ & Band & $0.45 \pm 0.13$ & $2.22 \pm 1.22$ & $266 \pm 37$ & $(6.32 \pm 0.12) \times 10^{-6}$ & & \\
     081128 & $102.3 \pm 10.6$ & Band & $0.36 \pm 0.77$ & $2.39 \pm 0.24$ & $42.3 \pm 24.9$ & $(2.04 \pm 0.18) \times10^{-6}$ & $39.3$ & 54 \\
     081211B$^b$ & $64.0 \pm 1.6$ & PL & $1.64^{+0.20}_{-0.19}$ & & $> 150$ & $(4.34^{+0.55}_{-0.53}) \times 10^{-7}$ & & \\
     090113 & $9.1 \pm 0.9$ & PL & $1.59 \pm 0.06$ & & $> 150$ & $(6.75 \pm 0.11) \times 10^{-7}$ & $43.1$ & 56 \\
     091029 & $39.2 \pm 5.0$ & Cutoff & $1.60 \pm 0.18$ & & $135.1 \pm 63.2$ & $(2.03 \pm 0.02)\times10^{-6}$ & $47.8$ & 55 \\
     131229A$^a$ & $13.0 \pm 0.2$ & Band & $0.73 \pm 0.02$ & $4.31 \pm 3.09$ & $379 \pm 13$ & $(2.64 \pm 0.01)\times10^{-5}$ & & \\
     180204A$^a$ & $1.16 \pm 0.12$ & Band & $0.88 \pm 0.06$ & $2.40 \pm 0.38$ & $814.8 \pm 164.9$ & $(1.75 \pm 0.01) \times 10^{-6}$ & & \\
     180614A & $7.1 \pm 0.9$ & PL & $1.51 \pm 0.16$ & & $> 150$ & $(1.91 \pm 0.08) \times10^{-7}$ & $71.0$ & 56 \\
     191101A & $142.7 \pm 18.2$ & PL & $1.64 \pm 0.12$ & & $> 150$ & $(1.72 \pm 0.05)\times10^{-6}$ & $41.5$ & 56 \\
     200215A$^c$ & $24.3 \pm 2.0$ & Cutoff & $0.76 \pm 0.16$ & & $162 \pm 25$ & $(1.86 \pm 0.18)\times10^{-6}$ & & \\
     210514A$^d$ & $\sim74.2$ & Band & $0.38 \pm 0.64$ & $2.27 \pm 0.33$ & $118 \pm 32$ & $(1.97 \pm 0.50)\times10^{-5}$ & & \\
     231230A & $15.2 \pm 0.9$ & PL & $1.26 \pm 0.06$ & & $> 150$ & $(1.96 \pm 0.04)\times10^{-6}$ & $47.8$ & 56 \\
    \hline
    \hline
    \end{tabular}
\end{table*}

Figure~\ref{fig:Ep_Eiso} shows our results for $E_{p,i}$ and $E_{\rm iso}$ against the sample presented in \citet{Minaev20}. 
We overlay the $E_{p,i}$ -- $E_{\rm iso}$ correlation from \citet{Minaev20}, using the Type I and Type II subsample results for short and long GRBs, respectively, and the fit parameters obtained with the means of the \citet{York04} and \citet{Deming11} approximation methods \citep[the final two columns of Table 3 in][]{Minaev20}.

\begin{figure}
    \centering
    \includegraphics[width=\linewidth]{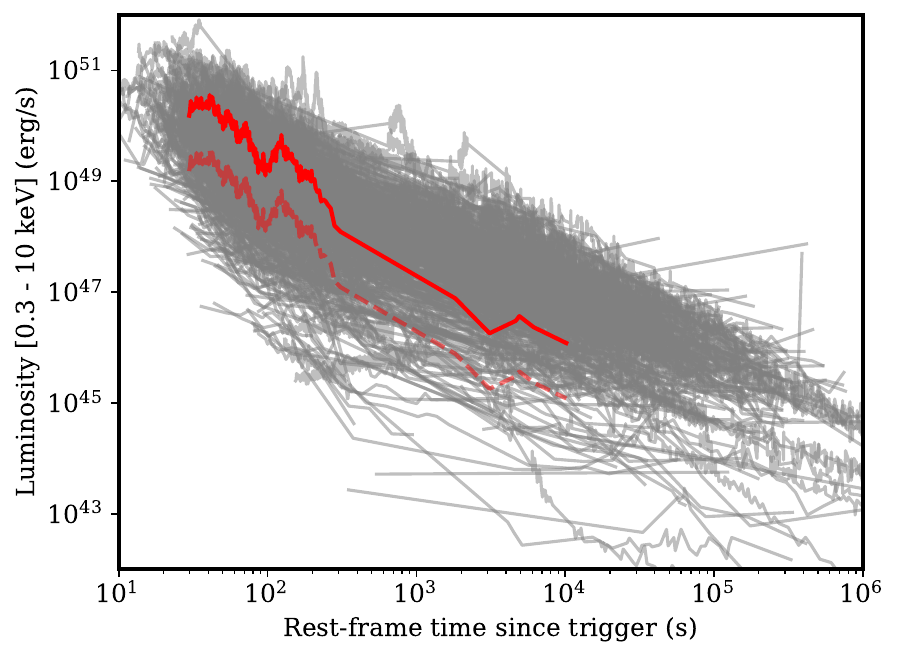}
    \caption{The X-ray afterglow of GRB 071031 (red) compared to X-ray afterglows of GRBs with known redshifts, taken from the UKSSDC. A k-correction based on the late-time XRT spectral fit has been applied to account for the shifting rest-frame bandpass at different redshifts. GRB 071031 appears unremarkable in luminosity space either as observed (solid red line) or assuming a magnification of $\mu = 10$, corresponding to an intrinsic luminosity ten times fainter (faint red dashed line). The X-ray afterglow is therefore uninformative with regards to the presence of lensing.}
    \label{fig:xray_comp}
\end{figure}

We first consider the 4 GRBs of known redshift, all of which are higher redshift than their cross-matched cluster. 
Three of them (GRB~060306, GRB~070419B and GRB~091029) are consistent with the $E_{p,i}-E_{\rm iso}$ relation within their measured uncertainties. This consistency indicates the magnification imparted by the nearby cluster is $\mu < 10$. In contrast, GRB~071031 shows a significant excess in $E_{\rm iso}$ relative to the constraint placed on its $E_{p,i}$ by spectral fitting. This is consistent with the expectations for a highly magnified ($\mu > 10$) source outlined above.
However, the large magnification required to explain the offset of GRB~071031 from the relation solely with lensing would imply the formation of multiple images, and we do not find any separate associated GRB detections (although, it is plausible they were missed).
In addition, GRB~071031's apparently anomalous $E_{\rm iso}$ is contingent on our $E_{p,i}$ estimate. Since the slope of the power law fit for this burst is $\alpha = 2.28\pm0.18$, it is within $2\sigma$ of the $\alpha=2$ threshold that would switch $E_{p}$ from an estimate of $< 15$keV to one of $> 150$keV using our definition based on the BAT bandpasses. If we instead consider the case of $E_{p} > 150$keV for this burst, $E_{\rm iso}$ is consistent with the long GRB Amati relation. Thus, based on our analysis, it is unclear whether GRB~071031 is highly magnified ($\mu > 10$), or is similar to the other three known-redshift GRBs ($\mu < 10$), but we slightly favour the $\mu < 10$ interpretation for the reasons described. We also compare the luminosity of GRB 071031's X-ray afterglow to the sample of GRBs with known redshifts on the UKSSDC, but find that it is uninformative with regards to lensing; the afterglow is of a `standard' luminosity whether it is unlensed or lensed by $\mu \sim 10$ (corresponding to an intrinsic luminosity that is a factor of ten times fainter; Figure~\ref{fig:xray_comp}).

We now consider the 11 GRBs without spectroscopic redshifts that we have in place assigned $z_{\rm GRB}=z_{\rm Cl}$. Of the 10/11 that are long GRBs, nine of them are faint outliers from the $E_{p,i}-E_{\rm iso}$ relation for long GRBs, each at $>3\sigma$ significance (GRB~210514A is the only exception). This is a strong indication that $z_{\rm GRB}>z_{\rm Cl}$, and thus these GRBs are de facto gravitationally lensed, albeit with unknown gravitational magnifications. It is also worth noting that, despite being classified as a long GRB based on the canonical $t_{90}=2{\rm s}$ cut \citep{Kouveliotou93}, the measured $E_{p,i}$ and $E_{\rm iso}$ for GRB~200215A is consistent with the relation for short GRBs. Similarly, some of the other $t_{90}>2{\rm s}$ bursts with only lower limits on $E_{p,i}$ could be interpreted as a match to the short GRB relation when assigned the redshift of the nearby cluster. This may instead imply a merger-driven origin for these bursts and would explain some of the faint outliers from the long GRB relation seen here. This is important since the line between merger-driven and collapsar-driven GRBs has become blurred with the recent identification of merger GRBs with durations far in excess of the canonical 2s divide \citep{Rastinejad22,Mei22,Troja22,Yang22,Gillanders23,Gompertz23,Levan24,Yang24}. However, of our sub-sample, only GRB~081211B has been suggested to be a merger-driven event \citep{Golenetskii08}.

We also note that GRB~050509B, the only canonically-defined short GRB among our GRB-cluster matches 
%plotted in \autoref{fig:Ep_Eiso}, 
has only a lower limit on its $E_{p,i}$ value, which leaves open the possibility that it is under-luminous relative to $E_{p,i}$ even according to the short GRB $E_{p,i}-E_{\rm iso}$ \citep[see also][]{Gompertz20}. This could be interpreted as a mildly off-axis event, but it could also indicate the GRB originates from a higher redshift than the cluster and is therefore also lensed. This possibility is one of the reasons we highlight GRB~050509B in \autoref{sec:GRB050509B}, and therein present a new lens model of its nearby cluster.

\subsection{Refining cluster centres and GRB-cluster offsets}
\label{sec:bcg_refinements}

In general, lines of sight with smaller $\theta_{\rm sep}$ from a lens are subject to greater magnifications than those with larger $\theta_{\rm sep}$. In the absence of detailed lens models of the clusters, it is therefore important for $\theta_{\rm sep}$ to be calculated in a consistent manner across the sample of 15 GRB-cluster matches. This helps to strengthen evidence for lensing being at play, even in the absence of detailed models, and to correctly prioritize GRB-cluster matches for further detailed analysis and possible future follow-up observations. In that context, we note that the cluster centres were estimated in different ways for the respective cluster samples (Section~\ref{sec:cluster_sample}). Our generous $\theta_{\rm sep}<2'$ cut ensures no matches are excluded based on this difference, but the difference ultimately still affects the determined value of $\theta_{\rm sep}$.
We redefine the centres of clusters in our sample to all be based on the location of a known or candidate brightest cluster galaxy (BCG), for cases where they are not already, since the strong lensing region of a cluster tends to be centred on its BCG.

We use Legacy Survey and UHS/VHS imaging and colours alongside Legacy Survey photometric redshifts to search for obvious BCGs and other dominant galaxies in the respective cluster cores. We identified four clusters with a clear BCG within these images and $r-z$ red sequences that are offset from the cluster centre listed in the respective cluster catalogue sufficient to alter $\theta_{\rm sep}$ by more than 1 arcsec (see Appendix \ref{sec:LS_images} for Legacy Survey cutouts of these candidates).

We therefore refined our calculated $\theta_{\rm sep}$ for these cluster-GRB matches. The BCG coordinates, along with updated $\theta_{\rm sep}$ between the GRB and BCG are listed in \autoref{tab:updated_seps}. For all other clusters, either no clear red sequence was identified in the Legacy Survey data, or the cluster coordinates in \autoref{tab:matches} were verified as sufficient. 

All of the updated $\theta_{\rm sep}$ values in \autoref{tab:updated_seps} are smaller than the values in \autoref{tab:matches}. We discuss each in turn below.

\smallskip

\noindent\underline{GRB~091029}: Separation reduced from 22.49 arcsec to 8.77 arcsec. It remains the GRB-cluster match with the smallest $\theta_{\rm sep}$, although its consistency with the Amati relation implies its magnification is not large enough to show as an outlier on top of the spread of the relation.

\smallskip

\noindent\underline{GRB~071031}: $\theta_{\rm sep}$ reduced from 1.00 arcmin to 35.28 arcsec. This separation indicates magnification suffered by the GRB will be modest, and not alone enough to explain any deviation from the Amati relation. 

\noindent\underline{GRB~200215A}: $\theta_{\rm sep}$ reduced from 1.33 arcmin to 58.56 arcsec. This separation indicates magnification suffered by the GRB will be modest, and not alone enough to explain any deviation from the Amati relation. 

\smallskip

\noindent\underline{GRB~050509B}: $\theta_{\rm sep}$ reduced from 1.65 arcmin to 12.66 arcsec.  

This significant reduction occurs due to the bimodal structure of ZwCl\,1234.0$+$02916 (redMaPPer 11161 in our sample), where the two massive galaxy clumps that underpin this distribution are separated by almost 2 arcminutes. The coordinates of ZwCl\,1234.0$+$02916/redMaPPer\,11161 from \autoref{tab:matches} are centred on the Eastern clump, but the brightest galaxy in Legacy Survey $z$-band is part of the Western clump closer to the GRB, hence the change in \autoref{tab:updated_seps}. (These are mass concentrations ``A'' and ``B'', respectively, in  \autoref{fig:GRB050509B}.) GRB~050509B is a promising lensed GRB candidate, with its relatively small separation to a massive cluster member, and being a possible outlier from the Amati relation if assigned the cluster redshift in \autoref{fig:Ep_Eiso}, given its $E_{p,i}$ is a lower bound. This, along with additional information related to the cluster \citep{Dahle13} published since the analysis of \citet{Pedersen05} prompts a more detailed look at GRB~050509B, which we present in the next section.

\begin{table*}
    \centering
    \caption{Updated values of $\theta_{{\rm sep}}$ and cluster coordinates using locations of BCG candidates.}
    \label{tab:updated_seps}
    \begin{tabular}{llllll}
        \hline
        GRB & GRB $\alpha, \delta$ & Err$_{90}$ $(\arcsec)$ & Galaxy Cluster ID & BCG $\alpha, \delta$ & $\theta_{\text{sep}}(\arcsec)$ \\
        \hline
        091029 & 60.1776, -55.9554 & 1.4 & CALICO\_S19855 & 60.1819, -55.9556 & 8.77 \\
        050509B & 189.0574, 28.9843 & 5.4 & redMaPPer\_11161 & 189.0537, 28.9830 & 12.66 \\
        071031 & 6.4058, -58.0592 & 1.5 & CALICO\_S41090 & 6.4023, -58.0496 & 35.28 \\
        200215A & 34.0794, +12.7710 & 1.4 & CALICO\_N856 & 34.0736, 12.7557 & 58.56 \\
        \hline
    \end{tabular}
\end{table*}

\section{Gravitational lens model for GRB~050509B}\label{sec:GRB050509B}

\begin{figure*}
    \centering
    \includegraphics[height=60mm]{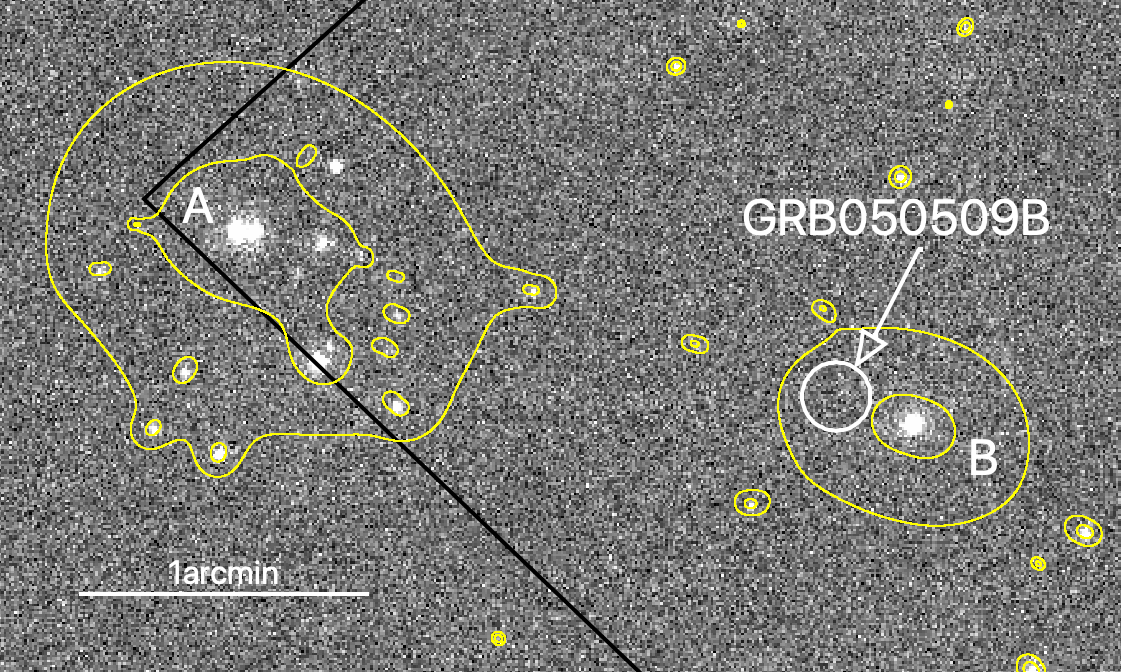}
    \hspace{2mm}
    \includegraphics[height=60mm]{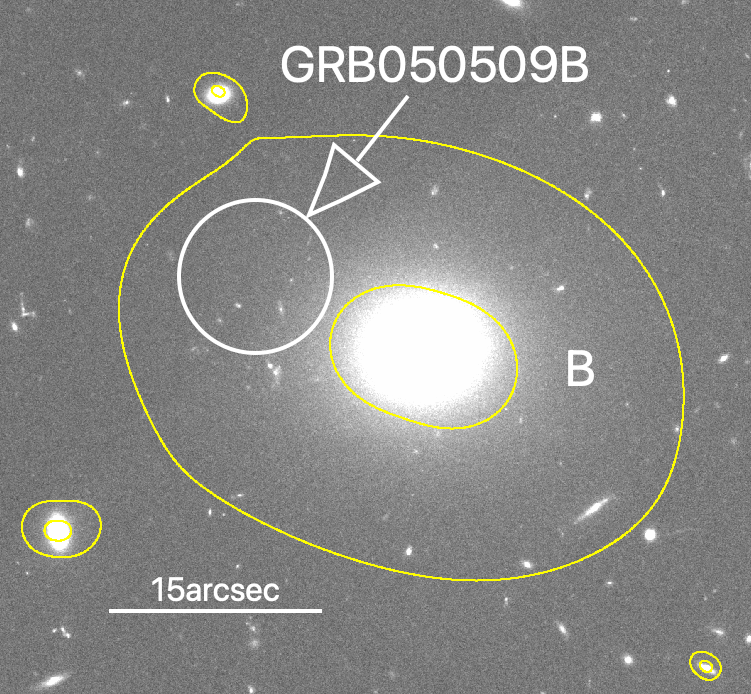}
    \caption{The bi-modal core of ZwCl\,1234.0$+$02916, showing that if GRB~050509B is located behind this cluster, then it is plausibly gravitationally magnified by $\mu\simeq2-6$, as described in Section~\ref{sec:GRB050509B}. \textsc{Left}: $J$-band imaging data from the UKIRT Hemisphere Survey are shown as the greyscale, lens magnifications of $\mu=2.5$ and $\mu=6$ external to the $z_{S}=2$ critical curves are shown as the outer and inner yellow contours respectively, based on the lens model in Section~\ref{sec:GRB050509B}. Rightward of the black line was included within the field of view of \emph{Hubble Space Telescope} imaging of this cluster/GRB with the Advanced Camera for Surveys through the F814W filter (Program ID: 10119; PI: D. Fox). \textsc{Right}: Zoom in to component B (at right in the left panel), showing the \emph{HST}/ACS data as the greyscale. All other details are as per the left panel. We note several candidate host galaxies of GRB~050509B residing within the positional error circle, all of unknown redshift. North is up and East is left in both panels.}
    \label{fig:GRB050509B}
\end{figure*}

\citet{Pedersen05} considered the possibility that GRB~050509B is located behind ZwCl\,1234.0$+$02916 and thus gravitationally lensed. They concluded that if GRB~050509B is lensed, then it is probably magnified by no more than a factor of 2, i.e.\ $\mu<2$. At that time, the error circle on the GRB afterglow localisation had a radius of $\theta=9.3''$ and was centered $9.8''$ from a massive early-type galaxy at a redshift of $z=0.225$ \citep{Gehrels2005}, and coincident with ``B'' in Fig.~\ref{fig:GRB050509B}. \citeauthor{Pedersen05}'s lens model comprised two mass components, one of which was centred on this massive galaxy, modelled as a singular isothermal ellipsoid with velocity dispersion $\sigma=260\pm40\,\rm km\,s^{-1}$ \citep{Bloom06}. The second component was a singular isothermal sphere of mass $M_{500}\simeq2\times10^{14}\rm M_\odot$ centred on the centroid of the X-ray emission from the galaxy cluster ZwCl\,1234.0$+$02916, where $M_{500}$ is the mass enclosed by the radius $r_{500}$ within which the mean density of the cluster is $500\times$ the critical density of the universe. The X-ray centroid is $\simeq20\,\rm arcsec$ West of ``A'' in Fig.~\ref{fig:GRB050509B} in the direction of ``B''. 

\citeauthor{Pedersen05} noted that their magnification estimate was dominated by the massive galaxy in their lens model, boosted slightly by the cluster component to the East. In other words, the sky position of GRB~050509B resides in the saddle region between two mass concentrations that they modelled as a massive cluster core and an individual early-type galaxy -- i.e.\ a bimodal mass distribution of unequal mass ratio. The enhanced lens magnification in saddle regions is well understood, because the density profile is flatter in a saddle region than in the absence of the second mass concentration, and lens magnification scales inversely with density profile slope \citep{Smith2025}. Specifically, the lens magnification suffered by sources lensed by clusters typically scales as $\mu\propto\eta^{-1}(2+\eta)^{-0.5}$, where $\eta$ is the slope of the density profile at the relevant location on the critical curve of the cluster lens (see \citeauthor{Smith2025} and references therein for details). 

Subsequent weak-lensing analysis \citep{Dahle13} probed the underlying mass and structure of the core of ZwCl\,1234.0$+$02916, relying on the sensitivity of lensing to all matter in the lens, not just the luminous matter. \citeauthor{Dahle13} identified that ZwCl\,1234.0$+$02916 has a bi-modal structure, with the two mass concentrations centred on the locations marked ``A'' and ``B'' in Fig.~\ref{fig:GRB050509B}. These two ``cluster-scale'' mass concentrations were found to have comparable masses, reminiscent of a ``Bullet-like'' cluster: $M_{\rm A}(<100\,{\rm kpc})=2.1\pm1.3\times10^{13}\rm M_\odot$, $M_{\rm B}(<100\,\rm {kpc})=1.3\pm0.3\times10^{13}\rm M_\odot$. \citeauthor{Dahle13} also found no evidence to support a mass concentration centred on the X-ray emission. In summary, \citeauthor{Dahle13} found that the mass ratio of the objects responsible for creating the saddle region behind which GRB~050509B may reside is much closer to unity than in \citeauthor{Pedersen05}'s model, and the mass concentration at ``B'' -- i.e.\ closer to GRB~050509B -- is a cluster-scale object not a galaxy-scale object. 

%Later weak-lensing analysis revealed that ZwCl\,1234.0$+$02916 has a bi-modal structure, with the two mass concentrations centred on mass concentration (hereafter ``A'') centred on massive early-type galaxies located $\simeq2\,\rm arcmin$ East of GRB~050509B, and one (hereafter ``B'') centred on the galaxy $\simeq9''$ West of GRB~050509B \citep{Dahle13}. These two ``cluster-scale'' mass concentrations were found to have comparable masses, reminiscent of a ``Bullet-like'' cluster: $M_{\rm A}(<100\,{\rm kpc})=2.1\pm1.3\times10^{13}\rm M_\odot$, $M_{\rm B}(<100\,\rm {kpc})=1.3\pm0.3\times10^{13}\rm M_\odot$. \citeauthor{Dahle13} also found no evidence to support a mass concentration centred on the X-ray emission. 

We therefore investigate whether \citeauthor{Dahle13}'s results imply a revision of \citeauthor{Pedersen05}'s gravitational magnification estimates. We use \textsc{LENSTOOL} \citep{Kneib1996,Jullo2007,Jullo2009} to model the core of ZwCl\,1234.0$+$02916 as two cluster-scale mass components centred on ``A'' and ``B'', following well-established methods. This includes describing the cluster-scale mass components as smoothly truncated pseudo-isothermal elliptical mass distributions \citep[PIEMD;][]{Kneib1996}, for which the projected density profile as a function of projected radius, $R$, is given by
\begin{equation}
    \Sigma(R)=\frac{\sigma_0^2\,r_{\rm cut}}{2G(r_{\rm cut}-r_{\rm core})}\left[\left(r_{\rm core}^2+R^2\right)^{-0.5}-\left(r_{\rm cut}^2+R^2\right)^{-0.5}\right],
\end{equation}
where $\sigma_0$ is the central velocity dispersion for a circular potential, $r_{\rm core}$ is the core radius, and $r_{\rm cut}$ is the cut-off radius  \citep{Limousin2005}. Note, we only consider circular potentials because no information is available to constrain ellipticity. For this model, the projected mass within a projected radius $R$ is given by 
\begin{equation}
    M(<R)=\frac{\pi\,r_{\rm cut}\sigma_0^2}{G}\left[1-\frac{\left(r_{\rm cut}^2+R^2\right)^{0.5}-\left(r_{\rm core}^2+R^2\right)^{0.5}}{r_{\rm cut}-r_{\rm core}}\right].
    \label{eqn:Maper}
\end{equation}
We adopt core and cut-off radii that are consistent with the literature, $r_{\rm core}=75\,\rm kpc$ and $r_{\rm cut}=1\,{\rm Mpc}$ \cite[e.g.][]{Richard2010,Fox2022}, noting that our results are insensitive to these choices. We use Equation~\ref{eqn:Maper} to obtain the following relationship between $M(<100{\rm kpc})$ -- that was measured for each cluster-scale mass using weak-lensing by \citeauthor{Dahle13} -- and central velocity dispersion, $\sigma_0$: 
\begin{equation}
    \sigma_0=520\,{\rm km\,s^{-1}}\left[{\frac{M(<100\,{\rm kpc})}{10^{13}\,\rm M_\odot}}\right]^{0.5}.
\end{equation}
We use this expression to estimate the velocity dispersion of each of the two cluster-scale mass components from $M_{\rm A}$ and $M_{\rm B}$, obtaining $\sigma_{0,\rm A}=750\,\rm km\,s^{-1}$ and $\sigma_{0,\rm B}=590\,\rm km\,s^{-1}$.  

We include galaxy-scale perturbers in the model, based on near-infrared photometry of cluster galaxies, following \cite{Smith2005} and many subsequent studies. This involves scaling the mass associated with each galaxy on its $K$-band Petrosian magnitude from UHS \citep{Dye18}, adopting the scaling relations first introduced by \citeauthor{Kneib1996}, and assigning an apparent $K$-band magnitude of $K=15.3$ to an $L^\star$ cluster galaxy at a redshift of $z=0.225$ \citep{Richard2010}. The massive galaxy adjacent to GRB~050509B and the massive galaxy in the centre of mass component A are assigned the velocity dispersion measured for the former by \cite{Bloom06}. We justify using the same velocity dispersion for both galaxies because their apparent magnitudes from the UHS differ by just $0.1$ magnitudes.

We use our lens model to estimate the lens magnification suffered by GRB~050509B in the scenario that it is located behind the cluster ZwCl\,1234.0$+$02916, that is at a redshift of $z_{\rm Cl}=0.225$. Placing GRB~050509B at a range of plausible GRB redshifts, $z_{\rm GRB}\simeq1-3$, we obtain $\mu\simeq2-6$, and show the corresponding magnification contours for $z_{\rm GRB}=2$ as an example in Fig.~\ref{fig:GRB050509B}. We also note that several faint galaxies are present within the GRB error circle, and are thus candidate host galaxies in the lensing interpretation that we have explored here -- see right panel of Fig.~\ref{fig:GRB050509B}. As far as we are aware, the redshifts of these galaxies have not yet been measured, and thus more refined magnification estimates are not currently possible. Moreover, the morphology of these galaxies does not show obvious signatures of strong lensing -- i.e.\ image multiplicity or gravitational shear -- in the single filter F814W \emph{HST}/ACS archival data presented in Fig.~\ref{fig:GRB050509B}. This underlines that our gravitational magnification estimate does not necessarily imply that GRB~050509B is strongly-lensed, i.e. multiply-imaged. If the lensing interpretation can be further explored and strengthened, for example with deep multi-band space-based observations and sensitive spectroscopic observations, then GRB~050509B may turn out to be in the singly-imaged regime analogous to gravitationally lensed supernovae that have been discovered at similar levels of magnification behind massive galaxy clusters \citep[e.g.][]{Goobar09,Amanullah11,Patel14,Rodney15,Rubin18}. This regime is typical of galaxy clusters due to their density profiles tending to be shallower on the scale of strong lensing than galaxy-scale lenses, rendering them inefficient at producing multiple images at  $\mu\simeq2-10$ \citep[][see also Fig.~\ref{fig:muredshift}]{Fox2022,Smith2023,Smith2025}. This would be consistent with the lack of time-separated repeated GRB detections, although it is possible that GRB signals associated additional images were too weak to detect. In general, it is possible to obtain further evidence for the lensing hypothesis post-hoc if a lensed host galaxy is found in deep, targeted imaging of the GRB localisation region, in addition to searching for other multiply-imaged background galaxies with which to constrain the lens model. This however, is beyond the scope of this paper, as it would require new observations with sensitive instruments on large ground- and space-based telescopes.

\section{Summary and discussion}
\label{sec:discussion}

In this study we have explored the possibility that some of the GRB detections discovered in the last $\sim 20$ years are lensed, due to proximity to a massive galaxy group or cluster. We cross-matched \emph{Swift} GRB detections with a sample of galaxy clusters from three main catalogues: CALICO, SDSS redMaPPer and WHY. We find 17 GRBs located within a 2 arcminute separation of a cluster within this sample. We looked further at 15 of the 17 GRB-cluster pairs, since either the GRB or the cluster has a redshift measurement or estimate. The 4 GRBs with confirmed redshifts are all behind their associated clusters, and hence will be affected by lensing to some degree. We investigate using the Amati relation between $E_{p,i}$ and $E_{\rm iso}$ to quantify the magnification: we find 3 of the 4 candidates are consistent with the Amati relation, implying their magnification is small, $\mu < 10$, whilst the other candidate, GRB~071031, is a significant ($>3\sigma$) outlier from the Amati relation. This would imply $\mu > 10$ for GRB~071031, but uncertainty in the obtained fit parameters means consistency with the Amati relation is also possible at the $2\sigma$ level. Therefore the conclusion of $\mu >$ or $< 10$ for GRB~071031 is unclear, with $\mu < 10$ gaining some credibility from no detection of multiple images. The luminosity of GRB 071031's X-ray afterglow is not a clear outlier from the wider X-ray afterglow population, and hence does not provide any further constraints on the scale of any possible magnification.

For the 11 GRBs without known redshifts, we used the Amati relation to explore whether these GRBs are consistent with being located at the cluster redshift. 10 of these candidates were shown to be under-luminous in $E_{\rm iso}$ with respect to the Amati relation when placed at the cluster redshift, indicating they are behind the cluster and hence lensed. Due to the separations between the GRB localisations and the cluster cores, the magnifications experienced by these GRBs will likely be modest. A possible exception is GRB~050509B, since it is very close to a massive galaxy within a merging cluster. We produce a gravitational lens model for this cluster, which predicts a magnification $\mu \simeq 2-6$ for GRB~050509B. 

Our finding of 14 GRBs that are: 1) located nearby a galaxy cluster and 2) at or consistent with a higher redshift origin than the associated cluster, is evidence for lensing within the \emph{Swift} GRB sample. Our magnification estimates of $\mu < 10$ (or possibly $\mu > 10$ for one of the 14) are consistent with expectations that GRBs lensed by galaxy clusters will be dominated by singly-imaged, $\mu<10$ sources (see \autoref{fig:muredshift}), compared to highly magnified and multiply imaged sources lensed by galaxy clusters, and lower magnification multiple-imaging by galaxy-scale lenses. Whilst this conclusion would benefit from more precise estimates of magnification, obtaining these are difficult since the GRBs and any associated afterglows fade within a $\sim$week. However, additional lensing information can be found post-hoc for these $\sim$arcsecond-localised events. By utilising deep imaging of the GRB localisation, one can locate a GRB host galaxy lensed by the cluster. Such observations would also provide extra constraints on the cluster lens model, and hence GRB magnifications, through positions of multiply-imaged galaxies.

Having shown there is evidence for lensing of GRBs in an archival search, it is pertinent to consider how future live lensed GRB searches may look, in an effort to make the first spectroscopically-confirmed lensed GRB with resolved multiple images. Such lensed GRBs provide massive scientific value and can even be associated with a lensed gravitational wave (see \citealt{Andreoni2024,Smith2025,Levan2025} for details), but making the first detection will be challenging. In the case where, like in our \emph{Swift} sample, X-rays are detected alongside the gamma-ray emission, the localisation is significantly reduced from $\sim 1000$ deg$^2$ to a few arcseconds$^{2}$. One could then find candidates with a live cross-matching of GRB localisations to a watchlist of capable lenses -- similar in essence to this study -- and follow up the localisation region to look for a lensed optical counterpart and/or host galaxy with a single telescope pointing. Another useful tracer in this case is a significant $E_{\rm iso}$ outlier, which can be useful in case the lens is faint or otherwise unknown. $E_{\rm iso}$ outliers are not so useful for rapid follow-up, however, since the GRB redshift is required before its position on the $E_{p,i}$ - $E_{\rm iso}$ plane can be determined. Such a tracer would still be useful for post-hoc observations to locate the lensed host galaxy, and this will be particularly valuable if a second image (or more) are yet to arrive to prepare future observations.

It is important to note X-rays are only detected in conjunction with a GRB in $\sim 20$ per cent of cases. Therefore, in the majority of cases, follow-up has to cover $\sim 1000$ deg$^2$ GRB localisations, and these observations must be completed immediately to find the optical counterpart, since post-hoc observations will be fruitless over such a large region housing tens of thousands of possible lenses. As for trigger criteria in these cases, the rapid turn around requirement rules out $E_{\rm iso}$ outliers, and the large localisations rule out picking out a single possible nearby lens. Thus, we must rely on individual detections of multiple images with similar inferred properties (i.e. overlapping sky localisations), hinting they emerged from the same GRB. Time delays between lensed images vary widely from seconds to decades, depending on many factors including the type of lens (single galaxy or group/cluster) and the source-lens alignment. Lensed GRBs with short time delays (on the order of $t_{90}$) will manifest as a single multi-peaked light curve, and will be difficult to disentangle from non-lensed GRBs, of which many intrinsically have multiple peaks. Lensed GRBs with longer time delays (of multiple years) will also be harder to discover this way, since a longer baseline opens up a higher probability for a false positive overlap. Despite this, rates of detectable lensed GRBs are expected to be $\sim 0.5$ per year with current \emph{Fermi} and \emph{Swift}  GRB satellites \citep{Andreoni2024, Smith2025}, and with arrival time differences of order months, i.e.\ comparable with lensed quasars used for cosmology. Furthermore, the feat of probing depths of $\lesssim 24$ mag and tiling the $1000$ deg$^2$ localisations in one night required for the majority of lensed GRBs is becoming feasible due to large field-of-view telescopes that dedicate a fraction of their time to target-of-opportunity observations, such as the Vera C. Rubin Observatory, the La Silla Schmidt Telescope (that will conduct the LS4 survey; \citealt{Miller2025}) and the Gravitational wave Optical Transient Observer (GOTO; \citealt{Steeghs2022}). Particularly Rubin, with its effective 6.4m mirror and 9.6 deg$^2$ field of view, is set to begin observations in early 2026 and is particularly well-placed to transform lensed GRB discoveries from a concept into a reality through its ToO programme \citep{Andreoni2024}.

Rubin's LSST will also provide a promising avenue to discover lensed GRBs through its routine survey, including so-called ``orphan afterglows''. Orphan GRB afterglows do not have a detectable associated gamma-ray counterpart, since their cone of gamma-ray emission is not oriented towards Earth, making them only discoverable through optical surveys like LSST or by X-ray telescopes. Studies predict LSST is capable of detecting around 50 un-lensed orphan afterglows per year \citep{Ghirlanda15, Lamb18}, and $\sim 0.05$ lensed orphan afterglows per year \citep{Gao22} -- a value which becomes interesting across the entire 10-year survey. Depending on the time delay between images of a lensed orphan afterglow, they could in principle be identified from LSST data alone. However, these very rare lensed events will be difficult to disentangle from other, more common fast-evolving optical transients such as FBOTs, stellar flares or AGN activity, and would likely require dedicated follow-up resources in order to confidently rule out false positives.

\section*{Acknowledgements}

DR acknowledges funding from the European Research Council (ERC) under the European Union’s Horizon 2020 research and innovation programme (LensEra: grant agreement No 945536).
B.P.J. acknowledges an Undergraduate Summer Bursary from the Royal Astronomical Society. 
BPG acknowledges support from STFC grant No. ST/Y002253/1
and The Leverhulme Trust grant No. RPG-2024-117. G.P.S. acknowledges support from The Royal Society, the Leverhulme Trust and the Science and Technology Facilities Council (grant number ST/X001296/1). 

The authors thank Tom Collett for enlightening discussions during the completion of this work.

The Legacy Surveys consist of three individual and complementary projects: the Dark Energy Camera Legacy Survey (DECaLS; Proposal ID 2014B-0404; PIs: David Schlegel and Arjun Dey), the Beijing-Arizona Sky Survey (BASS; NOAO Prop. ID 2015A-0801; PIs: Zhou Xu and Xiaohui Fan), and the Mayall z-band Legacy Survey (MzLS; Prop. ID 2016A-0453; PI: Arjun Dey). DECaLS, BASS and MzLS together include data obtained, respectively, at the Blanco telescope, Cerro Tololo Inter-American Observatory, NSF’s NOIRLab; the Bok telescope, Steward Observatory, University of Arizona; and the Mayall telescope, Kitt Peak National Observatory, NOIRLab. Pipeline processing and analyses of the data were supported by NOIRLab and the Lawrence Berkeley National Laboratory (LBNL). The Legacy Surveys project is honored to be permitted to conduct astronomical research on Iolkam Du’ag (Kitt Peak), a mountain with particular significance to the Tohono O’odham Nation. The Photometric Redshifts for the Legacy Surveys (PRLS) catalog used in this paper was produced thanks to funding from the U.S. Department of Energy Office of Science, Office of High Energy Physics via grant DE-SC0007914.

This work made use of data supplied by the UK Swift Science Data
Centre at the University of Leicester. 

\section*{Data Availability}
The data underlying this article were accessed from various online public sources, referenced in the relevant sections. The authors have endeavoured to describe analysis techniques as clearly as possible to ensure reproducibility and to appropriately reference publicly-available analysis packages where appropriate. The derived data generated in this research will be shared on reasonable request to the corresponding author.

%%%%%%%%%%%%%%%%%%%% REFERENCES %%%%%%%%%%%%%%%%%%

% The best way to enter references is to use BibTeX:

\bibliographystyle{mnras}
\bibliography{lensed_grbs_method} % if your bibtex file is called example.bib

% Alternatively you could enter them by hand, like this:
% This method is tedious and prone to error if you have lots of references
%\begin{thebibliography}{99}
%\bibitem[\protect\citeauthoryear{Author}{2012}]{Author2012}
%Author A.~N., 2013, Journal of Improbable Astronomy, 1, 1
%\bibitem[\protect\citeauthoryear{Others}{2013}]{Others2013}
%Others S., 2012, Journal of Interesting Stuff, 17, 198
%\end{thebibliography}

%%%%%%%%%%%%%%%%%%%%%%%%%%%%%%%%%%%%%%%%%%%%%%%%%%

%%%%%%%%%%%%%%%%% APPENDICES %%%%%%%%%%%%%%%%%%%%%

\appendix

\section{Estimating CALICO cluster redshifts}

\label{sec:calico_redshifts}
As described in \autoref{sec:cluster_sample}, CALICO does not determine a redshift estimate for any of the clusters in its sample. Therefore, for clusters with associated GRB cross-matches, it is imperative to determine a redshift estimate for use in our analysis. Where available, we can make use of the DESI (Dark Energy Spectroscopic Instrument) Photometric Legacy Survey, which provides photometric redshifts for galaxies within its almost $20,000$ square degree footprint. We estimate the photometric redshifts of CALICO clusters by first searching for an obvious candidate bright central galaxy (BCG), nearby to the CALICO coordinates -- these are typically the brightest of all cluster members and so will have the most reliable photometric redshifts. These are found by constructing colour-magnitude diagrams and identifying the brightest galaxy within a cluster red sequence. We use two diagrams for confidence, one using optical $r$ and $z$-band photometry from DESI Legacy Survey and another using NIR $J$ and $K$-band photometry from the UHS/VHS data used in CALICO. We also check for peaks in the photometric redshift distribution of the Legacy Survey data obtained within $2'$ from the cluster coordinates, and use these to substantiate cluster redshift estimates.

For 4 out of the 6 CALICO clusters cross-matched within $2'$ of a \textit{Swift}/XRT GRB, we were able to obtain photometric redshift estimates using the Legacy Survey data. These are quoted in \autoref{tab:matches}. 3 of the 4 are quoted with two redshift values, because there is evidence of two collections of objects at different redshifts within the field of the CALICO detection; one from a BCG identified by red sequence, and one from a peak in the photometric redshift distribution. It should be noted that typical photometric redshift errors $\Delta z$ on $z < 0.5$ galaxies in the fields of CALICO detections are fairly consistent at $\Delta z \lesssim 0.1$, implying the uncertainties are small enough to rule out these two redshift estimates instead originating from a single population of objects at a single redshift.

The remaining two CALICO clusters without redshift estimates do not fall in the Legacy Survey footprint, and independently do not show any strong evidence of a red sequence in the NIR photometry, despite a significant detection in CALICO. We therefore are unable to estimate their redshifts, and exclude these two GRB-cluster pairs from our analysis. Their associated GRBs do not appear in \autoref{fig:Ep_Eiso}.

\section{Images of cluster fields for select candidates}
\label{sec:LS_images}

\begin{figure}
    \centering
    \includegraphics[width=\columnwidth]{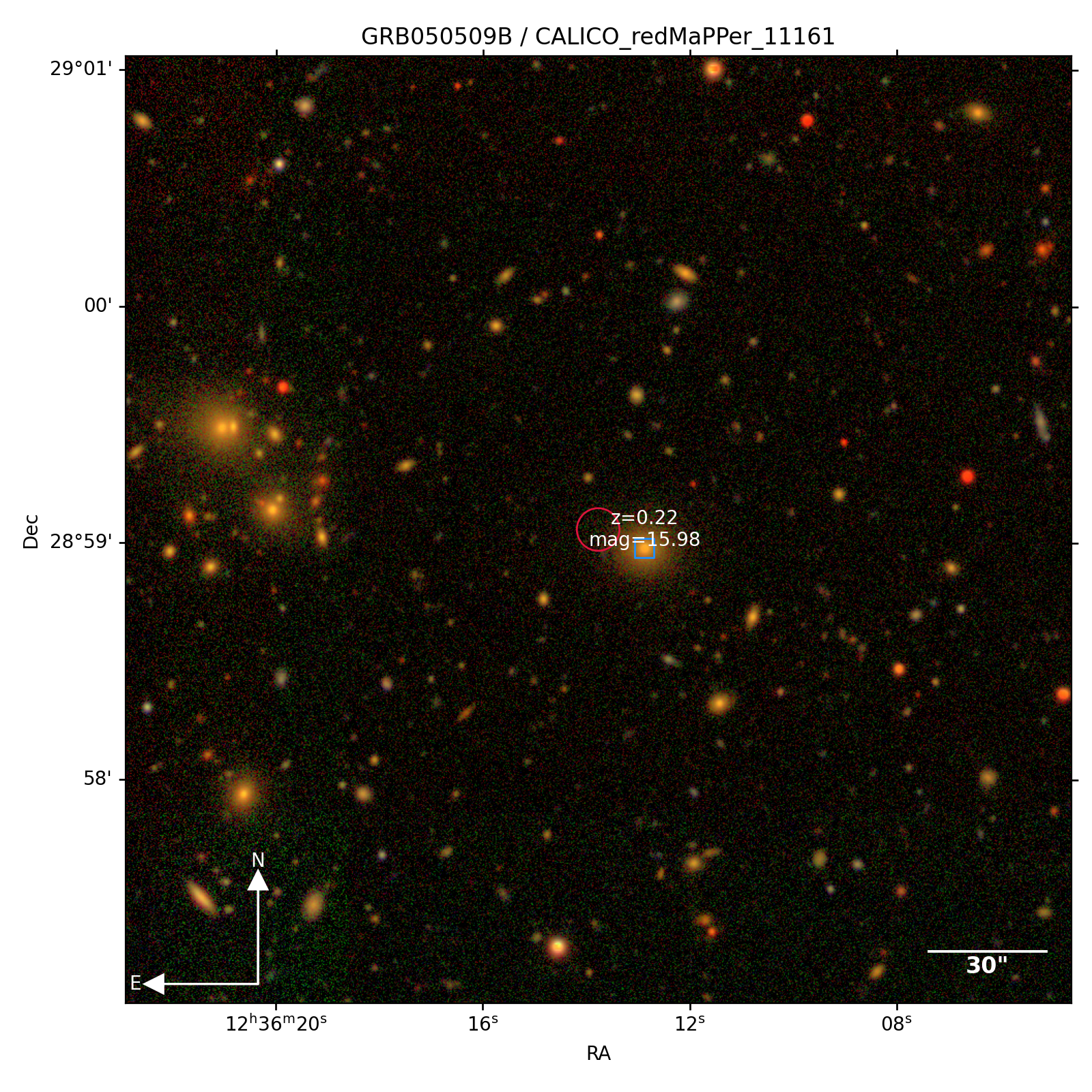}
    \caption{Legacy Survey RGB colour image ($gri$ bands) of the field containing GRB~050509B and redMaPPer 11161. The red circle shows the GRB localisation region, and the blue square highlights the BCG picked out with Legacy Survey and infra-red data, with its Legacy Survey photometric redshift and $z$-band magnitude shown.}
    \label{fig:LS_050509B}
\end{figure}

\begin{figure}
    \centering
    \includegraphics[width=\columnwidth]{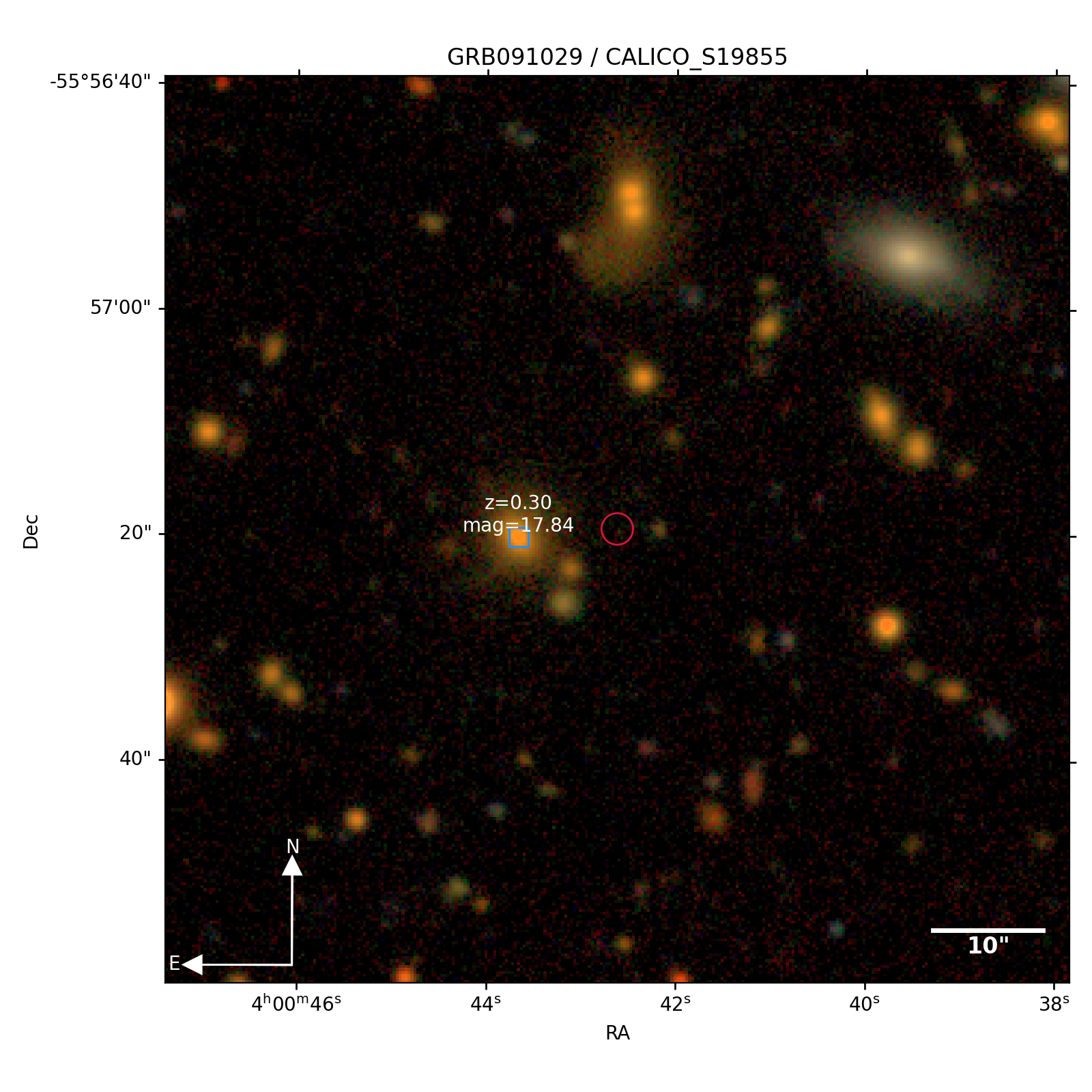}
    \caption{As for previous figure, but for the field of GRB~091029 and CALICO S19855.}
    \label{fig:LS_091029}
\end{figure}

\begin{figure}
    \centering
    \includegraphics[width=\columnwidth]{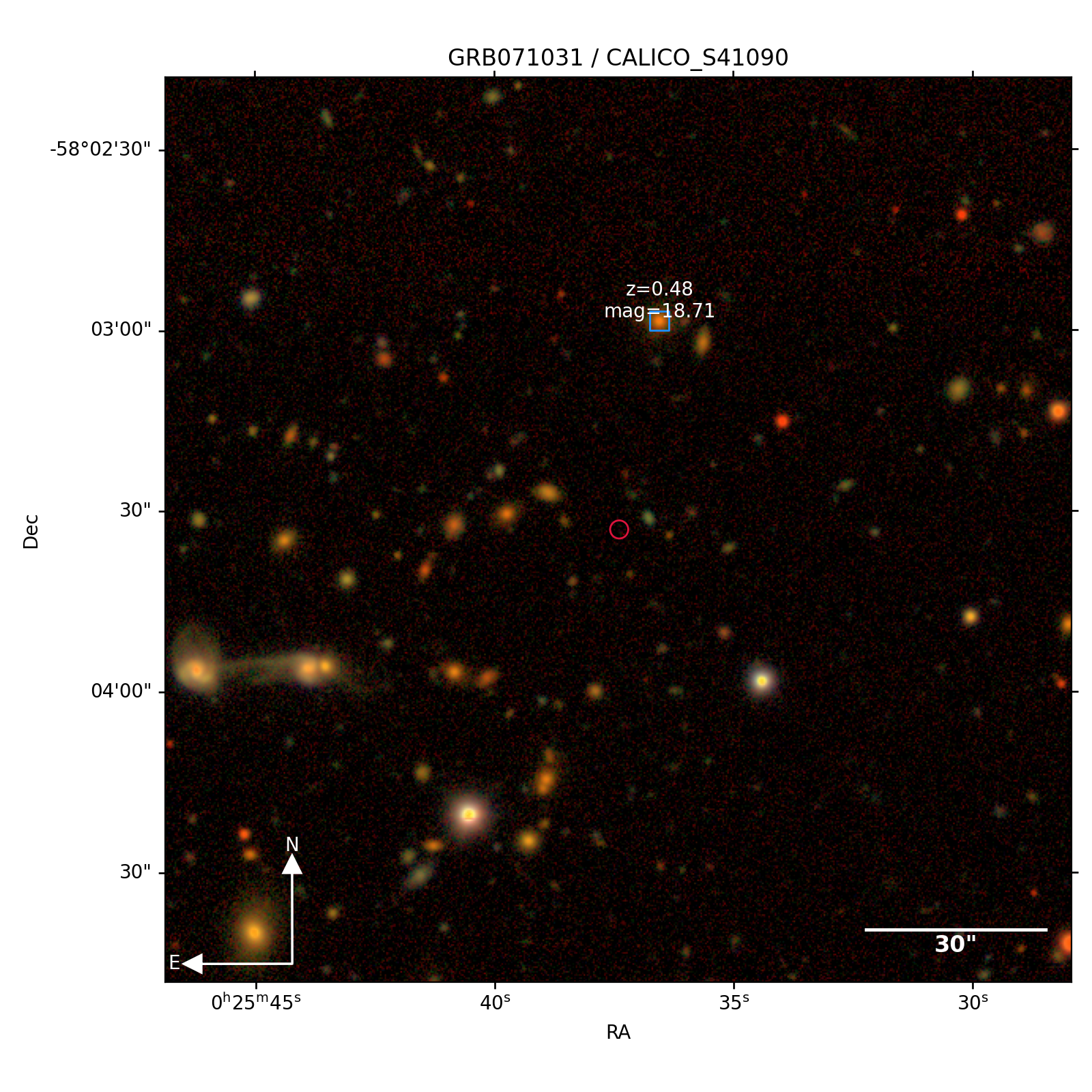}
    \caption{As for previous figure, but for the field of GRB~071031 and CALICO S41090.}
    \label{fig:LS_071031}
\end{figure}

\begin{figure}
    \centering
    \includegraphics[width=\columnwidth]{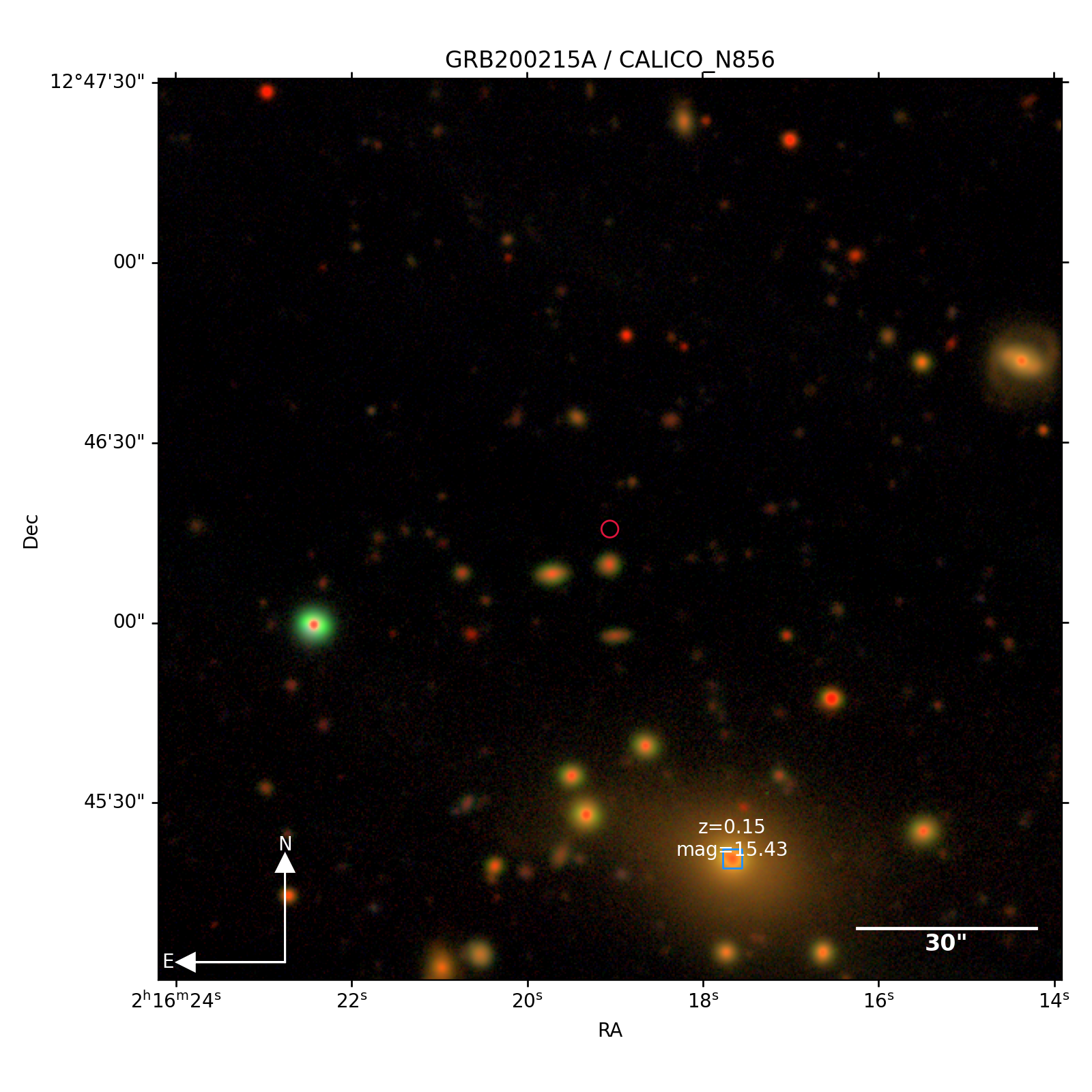}
    \caption{As for previous figure, but for the field of GRB~200215A and CALICO N856.}
    \label{fig:LS_200215A}
\end{figure}

%%%%%%%%%%%%%%%%%%%%%%%%%%%%%%%%%%%%%%%%%%%%%%%%%%

% Don't change these lines
\bsp	% typesetting comment
\label{lastpage}
\end{document}